\documentclass[10pt,
twocolumn,
 aip,
 jmp,%
 showkeys,
 amsmath,amssymb,
reprint,%
]{revtex4-1}

\usepackage[breaklinks=true,colorlinks=true,linkcolor=blue,urlcolor=blue,citecolor=blue]{hyperref}
\usepackage{graphicx}%
\usepackage{graphics}%
\usepackage{epstopdf}
\usepackage{dcolumn}%
\usepackage{soul,color}
\usepackage{bm}
\usepackage[mathlines]{lineno}
\expandafter\ifx\csname package@font\endcsname\relax\else
 \expandafter\expandafter
 \expandafter\usepackage
 \expandafter\expandafter
 \expandafter{\csname package@font\endcsname}%
\fi
\begin{document}


\title{Dual cut-off dc-tunable microwave low-pass filter on superconducting Nb microstrips with asymmetric nanogrooves}
\author{Oleksandr V.~Dobrovolskiy}
    \email{Dobrovolskiy@Physik.uni-frankfurt.de}
    \affiliation{Physikalisches Institut, Goethe University, 60438 Frankfurt am Main, Germany}
    \affiliation{Physics Department, V. Karazin National University, 61077 Kharkiv, Ukraine}
\author{Michael Huth}
    \affiliation{Physikalisches Institut, Goethe University, 60438 Frankfurt am Main, Germany}
\date{\today}

\begin{abstract}
We present a dual cut-off, dc-tunable low-pass microwave filter on a superconducting Nb microstrip with uniaxial asymmetric nanogrooves. The frequency response of the device was measured in the range $300$\,KHz to $14$\,GHz at different temperatures, magnetic fields, and dc current values. The microwave loss is most effectively reduced when the Abrikosov vortex lattice spatially matches the underlying washboard pinning landscape. The forward transmission coefficient $S_{21}(f)$ of the microstrip has a dc-tunable cut-off frequency $f_d$ which notably changes under dc bias reversal, due to the two different slope steepnesses of the pinning landscape. The device's operation principle relies upon a crossover from the weakly dissipative response of vortices at low frequencies when they are driven over the grooves, to the strongly dissipative response at high frequencies when the vortices are oscillating within one groove. The filter's cut-off frequency is the vortex depinning frequency tunable by the dc bias as it diminishes the pinning effect induced by the nanopattern. The reported results unveil an advanced microwave functionality of superconducting films with asymmetric (ratchet) pinning landscapes and are relevant for tuning the microwave loss in superconducting planar transmission lines.
\end{abstract}

\date{April 01, 2015}

\maketitle

Planar transmission lines made from superconducting thin films constitute elementary building blocks in today's circuit quantum electrodynamics~\cite{Wal04nat,Hof09nat} and advanced quantum information processing~\cite{Dic09nat}. In particular, they are the basis for superconducting qubits~\cite{Cla08nat}, resonators~\cite{Son09apl,Bot11apl,Cou13prb}, and various Josephson~\cite{Knu12pre} and Abrikosov~\cite{Wor09apl} fluxonic devices. For most of these applications, is desired the miniaturization of integrated circuits in conjunction with high clock frequencies, driven by the need to reduce the energy dissipation (losses) and improving the resonators' quality factors.

Superconducting circuit elements exploited in magnetic fields are known to suffer from losses due to motion of Abrikosov vortices~\cite{Zai07prb,Wor09apl}. Unless pinned, they increase noise and bit error rate in quantum interference devices (SQUIDs)~\cite{Koe99rmp}, raise unwanted vortex-assisted photon-count rates in detectors~\cite{Eng12prb}, and reduce quality factors~\cite{Bot11apl,Son09apl} and power handling capabilities~\cite{Lah02apl}. In addition, depinning of vortices has been shown to trigger flux avalanches close to the edges of coplanar waveguides (CPWs)~\cite{Awa11prb}. At the same time, several approaches to reduce microwave (mw) losses have been proposed. For the case of residual ambient fields it was shown that energy losses due to a small number of vortices, caught while cooling through the superconducting transition, can be reduced by trapping them within a slot patterned into the resonator~\cite{Son09apl}. For larger fields, antidots fabricated along the conductors' edges have been proven to effectively increase the quality factor of niobium CPW resonators~\cite{Bot11apl}. However, for circuit elements with a width $w\gtrsim100\,\mu$m these approaches cannot compete with patterning of the entire surface of the superconductor with periodic arrays of pinning sites~\cite{Wor09apl,Sol14prb,Jin10prb,Wor12prb}.

The dynamics of Abrikosov vortices at mw frequencies has been extensively studied both theoretically~\cite{Git66prl,Cof91prl,Bra91prl,Shk08prb,Shk11prb,Shk14pcm} and experimentally~\cite{Git66prl,Pom08prb,Zai07prb,Jin10prb,Wor12prb,Sol14prb,Bot11apl,Awa11prb,Wor09apl,Jan06prb,Sil11sst,Pom13snm,Lar15nsr}. The central notion in the theoretical treatment of the ac-driven vortex dynamics is the \emph{depinning frequency}~\cite{Git66prl,Pom08prb}. It is defined as the frequency separating the low-frequency regime where the pinning forces dominate and the vortex response is weakly dissipative, from the high-frequency regime where the frictional forces prevail and the response is strongly dissipative. On the experimental side, the mw-driven vortex dynamics has been studied in low-$T_c$\cite{Awa11prb,Jin10prb,Git66prl,Sol14prb,Jan06prb,Sil11sst,Pom13snm,Lar15nsr} and high-$T_c$~\cite{Zai07prb,Wor12prb,Pom08prb} superconductors. Among these, a notable part of the work was concerned with nanopatterned superconductors~\cite{Jin10prb,Wor12prb,Bot11apl,Sol14prb,Lar15nsr} for which advanced fluxonic device functionalities are anticipated and, under certain conditions, a stimulation of superconductivity by a GHz stimulus has been observed recently \cite{Lar15nsr}. This is because a larger diversity and a better control over the vortex dynamic regimes may be obtained therein by varying the thermodynamic quantities (temperature $T$, magnetic field $H$) and fine-tuning the driving parameters (dc current density $j$, mw power $P$ and frequency $f$).

In this Letter, we report the design and performance of a \emph{dual cut-off, dc-tunable low-pass microwave filter}. The device is made from a superconducting Nb microstrip in which the motion of Abrikosov vortices is strongly affected by a \emph{washboard} pinning potential induced by a nanogroove array. Due to the asymmetric steepness of the groove slopes, the depinning currents for the positive and the negative branches of the current-voltage curve (CVC) differ and this is why the filter's cut-off frequency, which is mediated by the vortex depinning processes, alters under dc bias reversal. In addition, we show that the mw loss is minimal for the fundamental spatial matching of the vortex lattice with the underlying pinning landscape.

\begin{figure}
\centering
    \includegraphics[width=1\linewidth]{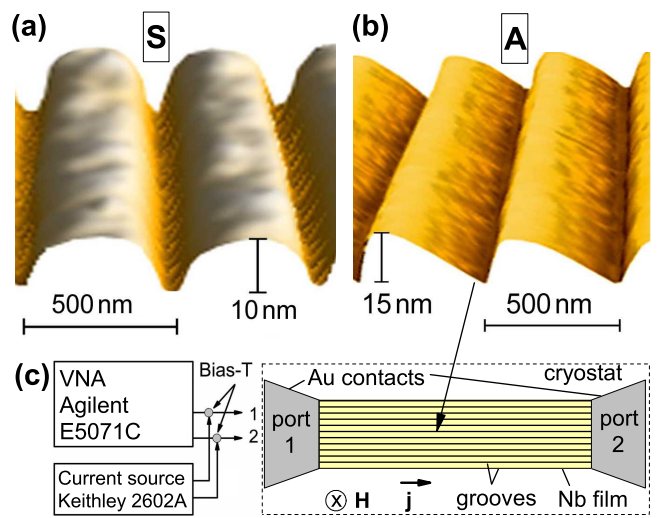}
    \caption{Atomic force microscope images of the nanogroove arrays fabricated on the surface of the Nb microstrips by FIB. Shown are grooves (a) with a symmetric cross-section in sample S and (b) with an asymmetric cross-section in sample A. (c) Block-diagram of the setup and the sample geometry.}
   \label{AFM}
\end{figure}

The $150\times500\,\mu$m$^2$ microstrips were prepared by photolithography from epitaxial (110) Nb films sputtered onto heated a-cut sapphire substrates~\cite{Dob12tsf}. The nanopatterns are uniaxial grooves with a groove-to-groove distance $a=500$\,nm, see Fig.~\ref{AFM}, fabricated by focused ion beam (FIB) milling~\cite{Dob12njp}. We note that the microstrip width is an integer multiple number ($N=300$) of the nanopattern period and the accuracy of FIB milling in conjunction with the large microstrip dimensions ensure that possible ratchet effects due to the edge barrier asymmetry \cite{Pry06apl,Ali09njp,Cer13njp} are insignificant to the highest attainable degree. The grooves are parallel to the direction of the transport current density $\mathbf{j}$ and, hence, in a perpendicular magnetic field $\mathbf{B}=\mu_0\mathbf{H}$ the dc dissipation and the microwave loss in the microstrip are related to the vortex dynamics \emph{across} the grooves, since they experience the action of the Lorentz force $\mathbf{F}_L = \mathbf{j} \times \mathbf{B}$ \cite{Sil10inb}. The data presented here were acquired on two microstrips, one patterned with grooves having a symmetric cross-section (sample S) and the other with grooves having an asymmetric one (sample A). In the patterning process this has been achieved by defining the grooves in the FIB bitmap file for sample S as a single line for the beam to pass, whereas a step-wise increasing number of FIB beam passes~\cite{Fib50apl} was assigned to each groove defined as a 5-step ``stair'' for sample A. In Fig.~\ref{AFM} one sees that due to blurring effects, the symmetric grooves in Sample S have rounded corners while smooth slopes resulted instead of the ``stairs'' in sample A. Samples S and A are 40\,nm and 70\,nm thick and have a critical temperature $T_c$ of $8.66$\,K and $8.94$\,K, respectively. The upper critical field for both samples at zero temperature $H_{c2}(0)$ is about $1$\,T as deduced from fitting the dependence $H_{c2}(T)$ to the phenomenological law $H_{c2}(T) = H_{c2}(0) [1-(T/T_c)^2]$.

Combined broadband mw and dc electrical measurements were done in a $^4$He cryostat in the temperature range $0.3T_c$ to $T_c$ with magnetic field $H$ directed perpendicular to the microstrip surface. A sketch of the experimental setup is shown in Fig.~\ref{AFM}(c) and we refer to Ref.~\cite{Dob15mst} for a detailed description of the custom-made sample probe. In brief, the sample was mounted in a copper housing in which pins of the micro-SMP connectors were spring-loaded against the pre-formed $300$~nm-thick gold contact pads sputtered through a shadow mask onto the film surface after the nanopatterning step. The mw signal was fed to the sample through coaxial cables from an Agilent E5071C vector network analyzer (VNA). The mw and dc signals were superimposed and uncoupled by using two bias-tees mounted at the VNA ports. The VNA operated in the frequency sweep mode, with 1548 frequency ($f$) points scanned with an exponentially growing increment between $300$\,KHz and $14$\,GHz. The frequency traces of the cables and the connectors were excluded from the data so that the recorded dependences are due to the microstrips themselves and the vortex dynamics therein. The measured quantity is the forward transmission coefficient $S_{21}$ being the ratio (expressed in dB) of the mw power measured at port $2$ to the power transmitted at port $1$. In what follows we use the notation $S_{21}$ for referring to its absolute value and primarily discuss the data for sample A. The data for the reference sample S are only commented when they are qualitatively different.
\begin{figure}
\centering
    \includegraphics[width=1\linewidth]{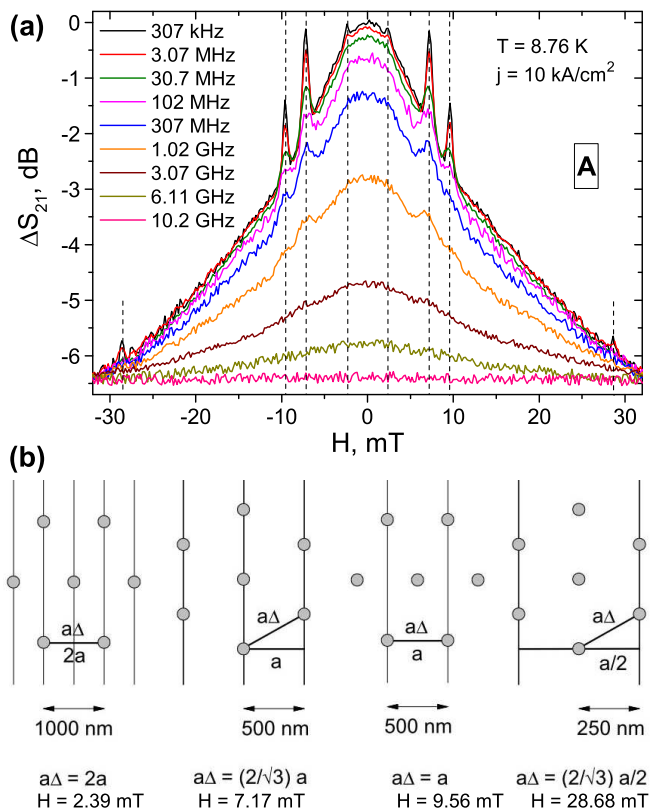}
    \caption{(a) Changes in the forward transmission coefficient $\Delta S_{21}$ of sample A under magnetic field reversal. (b) Vortex lattice configurations corresponding to the observed peaks in panel (a) along with the matching conditions $a_\bigtriangleup = ka$ and the calculated magnetic fields values.}
   \label{matching}
\end{figure}

Figure~\ref{matching}(a) presents the relative change of the forward transmission coefficient $\Delta S_{21} \equiv S_{21} - S_{21\mathrm{ref}}$ for sample A as a function of the magnetic field at $T = 0.98T_c$ and a series of frequencies, as indicated. Here, $S_{21\mathrm{ref}}$ stands for the mw loss in the transmission line (all cables, connectors etc.) and, hence, $\Delta S_{21}$ provides a measure for the mw loss due to vortex motion in the sample under study. For all frequencies, the mw excitation power at the sample is $P = - 20$\,dBm ($10\,\mu$W) as kept by the VNA in accordance with the pre-saved calibration data for $S_{21\mathrm{ref}}(f,T)$. The overall shape of the $\Delta S_{21}(H)$ curves in Fig~\ref{matching}(a) is symmetric with respect to $H=0$ and we admit that the close-to-critical temperature of the measurement might have masked a field-polarity dependence peculiar to asymmetric potentials~\cite{Plo09tas}. The data further attest to that at lower frequencies, the mw loss rises with increasing magnetic field, whereas $\Delta S_{21}$ becomes less sensitive to the field variation at elevated frequencies and saturates at the $-6.4$\,dB level (maximal loss). A detailed inspection of $\Delta S_{21}(H)$ unveils a pronounced reduction of the mw loss (peaks in $\Delta S_{21}$) at $7.2$\,mT and $9.6$\,mT, while less pronounced peaks (recognizable at $f\lesssim 3$\,MHz) correspond to $2.4$\,mT and $28.7$~mT. An increase of the microwave power by $20$\,dB leads to the disappearance of these matching field effects as the sample is driven to the normal state. Here we would like to note that commensurability effects were recently observed in Pb films with square pinning site arrays by magnetic field-dependent mw power reflection spectroscopy~\cite{Sol14prb} and broadband permeability transmission measurements \cite{Lar15nsr}. Also, the arrangement of vortices in a sawtooth, asymmetric pinning landscape was studied as a function of the vortex density by computer simulations \cite{Luq07prb}. While that system passed through a sequence of triangular, smectic, disordered and square vortex arrangements with increasing vortex density, the authors \cite{Luq07prb} outlined the presence of triangular ordering in the vortex arrangement corresponding to one column of vortices per pinning trough and argued that this vortex lattice configuration is the most stable one. In the present case, for the assumed triangular vortex lattice with lattice parameter $a_\bigtriangleup = (2\Phi_0/H\sqrt{3})^{1/2}$ and the geometrical matching conditions $a_\bigtriangleup = ka$, the arrangement of vortices with respect to the underlying pinning nanolandscape is shown in Fig.~\ref{matching}(b). The calculated field values for the triangular vortex lattices agree very well with those deduced from Fig.~\ref{matching}(a), although we cannot exclude the presence of more complex arrangements at fields 9.6\,mT and 28.7\,mT as suggested by simulations \cite{Luq07prb}. The field $7.2$\,mT corresponds to the maximal vortex density when each vortex is pinned at the bottom of a groove and there are no interstitial vortices, i.\,e. to the \emph{fundamental} matching configuration. It is this field for which the following findings are reported.

\begin{figure}
\centering
    \includegraphics[width=1\linewidth]{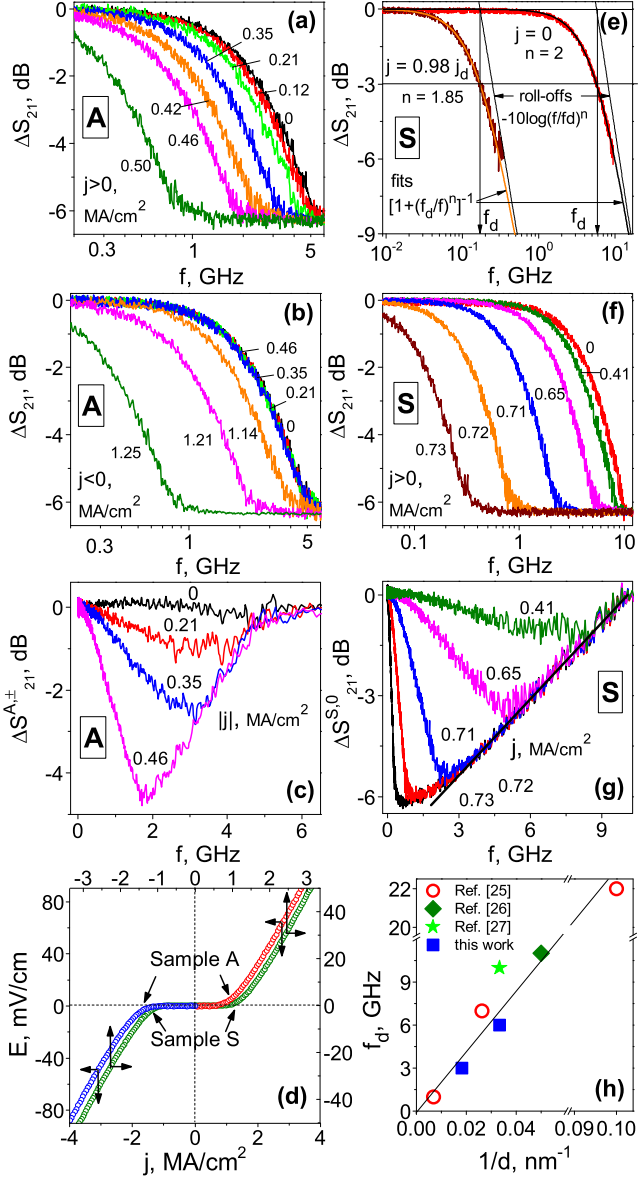}
     \caption{Dependence $\Delta S_{21}(f)$ of sample A at positive (a) and negative (b) dc current densities, as indicated. (c) Difference signal $\Delta S^{A,\pm}_{21}(f) = \Delta S_{21}(j) - \Delta S_{21}(-j)$ for sample A. (d) CVCs of sample A (left and bottom axes) and sample S (right and top axes) in the transition region from the pinned to the flux-flow regime. (e) Definition of the depinning frequency $f_d$ by the $-3$\,dB criterion exemplified for sample S along with fits of the filter's frequency characteristic to the expression $\Delta S_{21}(f) = 1/[1 + (f_{d}/f)^n]$. The filter roll-offs $-10\log_{10}[(f/f_{d})^n]$ are depicted by the straight lines with the exponents $n$ labeled close to the curves. (f) Frequency dependence of the forward transmission coefficient $S_{21}(f)$ of sample S for a series of dc densities, as indicated. (g) Difference signal $\Delta S^{S,0}_{21}(f) = \Delta S_{21}(j \neq 0) - \Delta S_{21}(j = 0)$ for sample S. In panels (a)-(g) $H = 7.2$\,mT, $T = 0.3T_c$, and $P = - 20$\,dBm. (h) Zero-bias depinning frequency at $H=0$ and $T=0$ as a function of the (inverse) Nb film thickness. The straight line $f_d \propto 1/d$ is guide for the eye.}
\label{power}
\end{figure}

Figure~\ref{power}(a) and (b) show the frequency dependence $\Delta S_{21}(f)$ of sample A for the same magnitudes of positive and negative dc current densities at $T = 0.3T_c$. In the absence of a dc bias, the mw loss is maximal at high frequencies, whereas the vortex response is weakly dissipative at low frequencies. For both dc bias polarities, upon increasing the dc value, the $\Delta S_{21}$ curves \emph{shift} towards lower frequencies, see Fig.~\ref{power}(a) and (b), but the magnitudes of the shifts \emph{substantially differ}, whereas these \emph{nearly coincide} for sample S (see supplementary figure \cite{Suppl}). A series of curves for the positive polarity for sample S is shown in Fig.~\ref{power}(f).

The reduction of the depinning frequency upon increasing the dc bias can be understood as a consequence of the effective lowering of the pinning potential well due to its tilt by the dc current. Indeed, the mechanistic consideration~\cite{Shk11prb} of a vortex as a particle leads to the conclusion that during an ac semiperiod, while the pinning potential well is broadening, with increasing $f$ the vortex has no longer time $(\sim 1/f)$ to reach the areas where the pinning forces dominate and, hence, the response becomes stronger dissipative already at lower frequencies, as compared with the zero-bias curve. The same mechanistic scenario can be applied for the explanation of the difference in the shifts of the depinning frequencies for the positive and the negative dc biases, caused by the different groove slope steepnesses~\cite{Shk14pcm}. Thus, by electrical resistance measurements we reveal that the groove asymmetry causes a difference in the depinning currents ($j_d$) for the positive and the negative branches of the CVC for sample A. At $T=0.3T_c$ and $H=7.2$\,mT these amount to $0.52$\,MA/cm$^2$ and $1.25$\,MA/cm$^2$, respectively. This is in contrast to the CVC of sample S which is symmetric with $j_d = 0.75$\,MA/cm$^2$, see Fig.~\ref{power}(d). Here, $j_d$ is determined by the $10\,\mu$V/cm electric field strength criterion. Therefore, sample A exhibits a \emph{microwave filter} behavior, whose \emph{cut-off frequency depends not only on the dc value} (as in sample S) \emph{but also on the dc polarity}. In other words, \emph{provided one changes the dc polarity while keeping the dc current absolute value, the device is switched from one cut-off frequency to another}.

To further corroborate our claim above, the difference signal $\Delta S^{A,\pm}_{21}(f) = \Delta S_{21}(j) - \Delta S_{21}(-j)$ for sample A is plotted in Fig.~\ref{power}(c). One clearly sees the difference between the mw loss measured for the \emph{same}, positive and negative dc bias magnitudes. This is caused by the fact that at moderate dc values (with respect to the corresponding $j_d$), for the gentle-slope direction of the vortex motion the depinning frequency is markedly lower than for the steep-slope direction. Specifically, the effect is most pronounced at $j = 0.46$\,MA/cm$^2$ at which the depinning frequencies are $f_d\approx1$\,GHz and $f_d\approx2.7\simeq3.02$\,GHz $= f_d(j=0)$ for the gentle-slope and the steep-slope direction, respectively. The depinning frequencies are determined at the $-3$\,dB loss level as shown in Fig.~\ref{power}(e). In addition, we note that $\Delta S^{A,\pm}_{21}$ looks qualitatively similar to the difference signal $\Delta S^{S,0}_{21}\equiv \Delta S_{21}(j\neq 0) - \Delta S_{21}(j = 0)$ for sample S in Fig.~\ref{power}(g). There, the data at higher frequencies fall onto a universal, nearly straight line reminiscent of the respective frequency dependence observed in high-$T_c$ films patterned with an array of antidots~\cite{Wor12prb}.

To quantitatively describe the observed filter behavior, we fit the $\Delta S_{21}(f)$ curves to the expression $\Delta S_{21}(f) = 1/[1 + (f_{d}/f)^n]$ with the exponent $n=2$ for $0 < j <0.42$\,MA/cm$^2$ and $n\approx1.85$ for $j\simeq j_d$ for sample S. For the gentle-slope direction of sample A $n\approx2.1$ and for its steep-slope direction $n\approx1.9$.  In addition, in both cases $n$ drops by about $10\%$ at close-to-depinning biases. In general~\cite{Poz11boo}, $n=2$ corresponds to a first-order filter roll-off $-10\log_{10}[(f/f_d)^2]$ related to the frequency characteristics of the microstrip as shown in Fig.~\ref{power}(e). The reduced filter effectiveness at $j\lesssim j_d$ is attributed to smearing of the crossover due to heating effects and a more pronounced role played by inhomogeneities of the current distribution at close-to-depinning dc current densities. We also stress that the mw filter operates under the fundamental matching field condition and the reported here effects become entangled and eventually vanish upon tuning the field value away from 7.2\,mT.

Finally, in Fig.~\ref{power}(h) we plot the zero-bias depinning frequencies at $T=0$ and $H = 0$, $f_d(0,0)$, for both samples along with the data for Nb films reported by other authors~\cite{Jan06prb,Sil11sst,Pom13snm}. The values $f_d(0,0)$ were estimated by the expression $f_d(T) = f_d(0)[1-(T/T_c)^4]$ successfully used~\cite{Zai03prb} for fitting the experimental data in high-$T_c$ films and the empiric dependence $f_d(H) = f_d(0)[1-(H/H_{c2})^2]$ observed~\cite{Jan06prb} for Nb films, respectively. For our data, the field and temperature correction of $f_d(T,H)$ to $f_d(0,0)$ amounts to about $1\%$, while it is notably larger for the data~\cite{Jan06prb,Sil11sst,Pom13snm} acquired at higher temperatures and larger fields. For our samples, their thicknesses were taken after subtracting the groove depths, see Fig.~\ref{AFM}. From Fig.~\ref{power}(h) it follows that the cumulative data for Nb films can be fitted to the phenomenological law $f_d \propto 1/d$, where $d$ is the film thickness.

In summary, we have observed two major effects in nanopatterned Nb microstrips by combined dc and microwave measurements. These effects are (i) microwave loss reduction due to spatial commensurability of the vortex lattice with the underlying washboard nanopattern and (ii) a dual cut-off, dc-tunable microwave low-pass filter behavior of the sample with uniaxial asymmetric nanogrooves \emph{under the fundamental matching field condition}. Since the effect of dc bias polarity-sensitive reduction of microwave losses may be exploited for tuning the properties of superconducting planar transmission lines, the reported results unveil an advanced microwave functionality of superconductors with an asymmetric (ratchet) pinning landscape. We furthermore anticipate that the observed filter behavior can be described by the results of Ref.~\cite{Shk14pcm}, while a detailed mapping of our findings on that model will be reported elsewhere.

OVD thanks Roland Sachser for automating the data acquisition and his help with the nanopatterning. Fruitful discussions with Valerij A. Shklovskij are gratefully acknowledged. This work was financially supported by the German Research Foundation (DFG) through grant DO 1511/2-4 and conducted within the framework of the NanoSC-COST Action MP1201 of the European Cooperation in Science and Technology.


\begin{thebibliography}{41}
\bibitem [{Fib()}]{Fib50apl}%
  \BibitemOpen
  \href@noop {} {}\bibinfo {note} {{In the patterning process the beam
  parameters were $30$\,kV/\,$50$\,pA, the dwell time was $1\,\mu$s, and the
  pitch was $50$\,nm}}\BibitemShut {NoStop}%
\end{thebibliography}
%

\end{document}